\documentclass[twocolumn]{svjour3}  
\smartqed  

\usepackage[dvipdfmx]{color}
\usepackage{graphicx}
\usepackage[T1]{fontenc}
\usepackage{mathptmx}  
\usepackage[scaled]{helvet}  
\usepackage{courier}
\usepackage[subrefformat=parens]{subcaption}

\makeatletter
\let\cl@chapter\undefined
\makeatletter
\usepackage{amsmath,amssymb}   
\usepackage{mathtools} 
\usepackage{cleveref}  
\Crefname{equation}{Eq.}{Eqs.}%
\Crefname{figure}{Fig.}{Figs.}%
\usepackage{amsbsy}
\usepackage{bm}
\usepackage{algorithmic}
\usepackage{algorithm}

\begin{document}

\title{Comparison of stochastic stability boundaries for parametrically forced systems with application to ship rolling motion
%
}

\author{Atsuo Maki         \and Yuuki Maruyama \and
        Yaliu Liu         \and Leo Dostal
}

\institute{Atsuo Maki \and Yuuki Maruyama \at
              Osaka University, 2-1 Yamadaoka, Suita, Osaka, Japan \\
              \email{maki@naoe.eng.osaka-u.ac.jp} 
           \and
           Leo Dostal \and Yaliu Liu \at
           Institute of Mechanics and Ocean Engineering, Hamburg University of Technology, 21043 Hamburg, Germany\\
           \email{dostal@tuhh.de} 
}

\date{Received: date / Accepted: date}

\maketitle

\begin{abstract}
%
Numerous accidents caused by parametric rolling have been reported on container ships and pure car carriers (PCCs). A number of theoretical studies have been performed to estimate the occurrence condition of parametric rolling in both regular and irregular seas. Some studies in random wave conditions have been the approximate extension of the occurrence conditions for regular waves (e.g. \cite{maki2011parametric}). Furthermore, several researches have been based on the stochastic process in ocean engineering (Roberts\cite{Roberts1982parametric} and Dostal\cite{dostal2012non}). This study tackled the parametric rolling in irregular seas from the stability of the system’s origin. It provided a novel theoretical explanation of the instability mechanism for two cases: white noise parametric excitation and colored noise parametric excitation. The authors then confirmed the usefulness of the previously provided formulae by Roberts and Dostal through numerical examples.

\end{abstract}

\section{Introduction}
    Parametric rolling, one of the threats to oceangoing vessels, is triggered by the change in the restoring moment caused by waves and from the viewpoint of nonlinear dynamical system theory, corresponds to parametric resonance. In the late 1990s, several accidents were reported on container vessels because of parametric rolling~\cite{france2003}, and since the 2000s, there have been also reports of it occurring at PCTCs~\cite{rosen2012experience}.
    
   In our paper survey, we have found Watanabe’s work~\cite{Watanabe1934} conducted in the 1930s. There is a long research history on parametric rolling, and theoretical studies have often been aimed at estimating the conditions of occurrence and motion amplitude of parametric rolling in regular waves. We present a review of representative studies to determine the conditions and amplitudes of parametric rolling in regular waves below.
    %
    Kerwin~\cite{kerwin1955} theoretically analyzed parametric rolling using the harmonic balance method with success, thus leading to subsequent works for predicting parametric rolling in regular seas by Zavodney et al.~\cite{zavodney1989}, Francescutto~\cite{francescutto2001}, Bulian~\cite{bulian2004approximate}, Spyrou\cite{spyrou2005paramet}, Umeda et al.~\cite{umeda2004nonlinear}, Maki et al.~\cite{maki2011parametric}, and Sakai et al.~\cite{sakai2018}.

    Research on parametric rolling in irregular waves has been very active in recent years, with a history that can be traced back to the 1980s~\cite{maki2011parametric,blocki1980,themelis2008}. Many studies have been conducted on this more generalized problem, taking a stochastic process approach in the fields of control engineering and mechanical engineering. For example, Samuels performed numerical experiments in which the damping term of the second-order differential equation was parametrically oscillated by white noise~\cite{samuels1960stability}. Caughey discussed this topic~\cite{caughey1960comments,samuels1960remarks}. Khasminskii~\cite{Khasminskii1967} discussed the stability of systems with parametric excitation terms, and Kozin~\cite{Kozin1971} further developed Khasminskii’s ideas, deriving stability conditions and comparing them with numerical calculations. For an overview of the progress made in these studies over time, references \cite{kozin1969} and \cite{Kozin1986} can be consulted.
    
     However, the problem we are addressing is not a system with a white noise parametric excitation term but one with a colored noise parametric excitation. For colored noise, obtaining results on the stability of the system’s origin becomes difficult, as the solution method used in the white noise case cannot be directly applied. To address this, Infante~\cite{Infante1968} obtained the condition of stability using the Lyapunov stability theory, while Arnold~\cite{Arnold1986Asymptotic} used perturbation expansions and calculated Lyapunov exponents to obtain analytical relations for stability.

    For systems with parametric excitation terms of the colored noise type, the problem can sometimes be solved using stochastic averaging methods. Stratonovich~\cite{Stratonovich1963} and Khasminskii~\cite{Khasminskii1966} developed the Stratonovich--Khasminskii limit theorem, which Roberts applied to ship problems~\cite{Roberts1982parametric}. More recently, Dostal proposed an energy-based stochastic averaging method~\cite{dostal2012non}, and Maruyama developed enhancements to this approach~\cite{maruyama2022improved}. Liu provides an extensive survey of the applicability of this method~\cite{liu2022applicability}. Maruyama~\cite{maruyama2021moment} also established an estimation method using the higher-order moment technique and successfully estimated the lateral angular acceleration using the results of this method~\cite{maruyama2021moment_accel}. Most of these studies focused on obtaining the PDF of the response, but some, such as the studies by Ariaratnam and Tam~\cite{Ariaratnam1979} and~\cite{dostal2012non}, refer to the stability of the origin of the system.

     This study aims to demonstrate the physics and mechanism of parametric rolling in irregular seas from the stability of the upright condition (the system's origin). In regular head sea cases, it is easy to understand that parametric rolling is attributed to the loss of asymptotic stability of the upright condition. For instance, as Maki et al.~\cite{maki2011parametric} demonstrated, its destabilization because of parametric excitation can be theoretically estimated by the averaging method. However, the occurrence of parametric rolling in irregular seas can be explained by the loss of stability of the system's origin. To explain this physics, the authors first present the system's stability with a stochastic coefficient modeled by the Wiener process. Secondly, they demonstrate the stability of the ship equation of motion with parametric excitation. In both considerations, several formulae to predict the stability threshold are examined. 

    Initial results from the investigation presented in this study were initially described by Maki et al. \cite{Maki2022parametric_JASNAOE}. In this paper, the results are presented more extensively, with more details, and with some revisions.

\section{Notations}\label{sec:notaions}
    In this study, the $n$-dimensional Euclidean space is denoted by $\mathbb{R}^n$, and the set of real numbers for $n=1$ is denoted by $\mathbb{R}$. The expectation operation is denoted by $\mathbb{E}$, and $t$ represents time. The overdot of time-dependent variables indicates the derivative with respect to time $t$.
    
\section{Equation of motion}\label{sec:state_equation}

    In this study, the authors deal with the single-DoF (Degree-of-Freedom) roll equation of motion:
    \begin{equation}
        \ddot{\phi} + 2 \zeta^{}\,\dot{\phi} + c_1 \phi + f(t)\,\phi = h(\,t\,).
        \label{1DoF_roll_equation}
    \end{equation}
    In this equation, $\phi(t)$ is the roll angle, $\dot{\phi}(t)$ is the roll angular velocity, and $\ddot{\phi}(t)$ is the roll angular acceleration. $2 \zeta$ is the damping coefficient, $c_1$ is the restoring moment coefficient, $f(t)$ is the parametric excitation term based on the restoring moment variation, and $h(t)$ is the roll moment caused by waves, each a function of time with a stochastic variation. In the colored noise case, the term $f(t)\phi$ is estimated using Grim’s effective wave concept~\cite{grim1961,umeda2022paramet}.
    
    This equation has linear damping and restoring components; thus, the nonlinear components are not essential for assessing the stability of the origin.
    
\section{Parametric oscillation for white noise}\label{sec:white_noise}
    %
    In this section, the authors briefly review the existing results conducted for the system with white noise parametric excitation. This is useful for understanding the physics of random parametric excitations, and therefore they consider the following parametric excitation term:
    \begin{equation}
        f(t) \phi(t) \mathrm{d}t = \Gamma \phi(t) \mathrm{d}W(t)
    \end{equation}
    Here, $\Gamma$ is a intensity strength of a white noise, and $W(t)$ is a 1D standard Wiener process that satisfies the relation
        \begin{equation}
            \left\{\begin{aligned}
                &\mathbb{E}[W(s) - W(t)] = 0\\
                &\mathbb{E}[(W(s) - W(t))^2] = |t-s|\\
                &~~~~~~~~~~~~\text{and}\\
                &\left\{\begin{aligned}
                    &\mathbb{E}[W(s)W(t)] = \min(s,t)\\
                    &~~~~~~~~~~~~~~~~\text{or}\\
                    &\mathbb{E}[\mathrm{d}W(s)\mathrm{d}W(t)] = \delta(t-s)
                \end{aligned}\right.
            \end{aligned}\right.
        \end{equation}
    and $\mathrm{d}W(t)$ is the increment of this Wiener process. Here, $\delta(t)$ means the Dirac's delta function. 
    \subsection{Some remarks on the stability in stochastic systems}\label{subsec:stability_white}
        In this subsection, we analyze the stability of the system origin with a parametric excitation term represented by white noise. We briefly explain the mathematical concepts used in this analysis.

        The problem addressed here is almost identical to that described in a recent paper by the authors, where the problem of ship maneuvering motion under white noise multiplicative noise from the aspect of stochastic disturbances is tackled\cite{Maki2022stabilization}. The governing equations, in that case, are almost identical to those in the present paper. However, attention must be paid to the presence of the Wong-Zakai’s correction term \cite{Wong-Zakai1965}, which appears when a real system is reduced to an Itô-type system of stochastic differential equations. If the Wong-Zakai’s correction term is present, it can strongly affect the stability of the system's origin, as discussed in \cite{kozin1969}. However, as mentioned in our previous work\cite{Maki2022stabilization}, the Wong-Zakai’s correction term does not exist in the system dealt with here.

        In the following subsections, the authors will demonstrate the method to identify the system's stability driven by white noise. First, the authors must carefully examine the relationship between the real system and the corresponding SDE to analyze the random system. The present system is the same as the one dealt with in our previous paper\cite{Maki2022stabilization}; hence, the Wong-Zakai’s correction term\cite{Wong-Zakai1965} is zero; this topic is discussed in the paper of Kozin\cite{kozin1969}.

        The roll angle and roll angular velocity are now represented as a state variable $x(t)\in \mathbb{R}^2$, hereafter.
        \begin{equation}
            x(t) \equiv \begin{bmatrix}
                x_1(t) \\
                x_2(t) \\
                \end{bmatrix} =
                \begin{bmatrix}
                \phi(t) \\
                \dot{\phi}(t) \\
                \end{bmatrix}
        \end{equation}
        Consider the following stochastic differential equation (SDE: Stochastic Differential Equation).
        \begin{equation}\label{eq:whiteNoiseParametricSDE}
            \mathrm{d}x(t) = \mu(x(t)) \mathrm{d} t + \sigma(x(t)) \mathrm{d} W(t)
        \end{equation}
        where $\mu(x(t)): \mathbb{R}^2 \rightarrow \mathbb{R}^2$, and $\sigma(x(t)): \mathbb{R}^2 \rightarrow \mathbb{R}^2$.

        Since there does not exist the Wong-Zakai’s correction term\cite{Wong-Zakai1965}, the drift and diffusion terms for the present system are given by $\mu(x(t),t)$ and $\sigma(x(t),t)$, respectively
        \begin{equation}
            \quad \left\{
            \begin{aligned}
                \mu(x(t),t) &= 
                \begin{bmatrix}
                x_2(t) \\
                - \kappa x_2(t) - c_1 x_1(t)
                \end{bmatrix}\\
                \sigma(x(t),t) &= 
                \begin{bmatrix}
                0 \\
                \Gamma x_1(t) \\
                \end{bmatrix},
            \end{aligned}
            \right.
        \end{equation}
        Finally, the SDE that the authors tackled is as follows:
        \begin{equation}
            \left\{ 
            \begin{aligned}
                \mathrm{d}x_1(t) &= x_2(t) \mathrm{d} t\\
                \mathrm{d}x_2(t) &= -(2 \zeta x_2(t) + c_1 x_1(t)) \mathrm{d} t + \Gamma x_1(t) \mathrm{d}W(t)\\
            \end{aligned}
            \right.
            \label{eq:system_white}
        \end{equation}

        We define the infinitesimal operator $\mathcal{L} \left[ \cdot \right]$ to analyze the dynamics of the functional $f(x(t)) \in \mathbb{R}$ of $x(t)$ as:
        \begin{equation}\left\{
            \begin{aligned}
                \mathcal{L} \left[ \cdot \right] \equiv & \left[ \frac{\partial (\cdot)} {\partial x} \right] \mu(x(t),t) \\ 
                & +\frac{1}{2} \sigma^T(x(t),t) \left[ \frac{\partial^2  (\cdot) }{\partial x^2} \right] \sigma(x(t),t).\\
                &\text{or}\\
                \mathcal{L} \left[ \cdot \right] \equiv &  \sum_{i=1}^2 \mu_i(x(t),t) \frac{\partial}{\partial x_i} \\
                & + \frac{1}{2} \sum_{i=1}^2 \sum_{j=1}^2 \left[ \sigma(x(t),t) \sigma^T(x(t),t) \right]_{ij} \frac{\partial}{\partial x_i \partial x_j}
            \end{aligned}\right.
            \label{eq:infinitesimal_operator_use}
        \end{equation}
        %
        For the system that the authors tackle, the infinitesimal operator for a scalar function $F(x(t)) \in \mathbb{R}$ becomes:
        \begin{equation}
            \begin{aligned}
                \mathcal{L} F(x(t)) = & x_2(t) \frac{\partial F(x(t))}{\partial x_1} \\
                & - (2\zeta x_2(t) + c_1 x_1(t)) \frac{\partial F(x(t))}{\partial x_2} \\
                &+ \frac{\Gamma^2}{2} x_1^2(t) \frac{\partial^2 F(x(t))}{\partial x_2^2}
            \end{aligned}
        \end{equation}
        The final term in this equation, $\Gamma^2/2 \cdot x_1^2(t) \cdot \partial^2 F(x) / \partial x_2^2$, is a characteristic term that appears in the framework of stochastic differential equations and can serve as a correction term for the dynamics of $F(x)$ because of the addition of noise. This term can either destabilize or stabilize the system.

        We consider the system which does not have the term of parametric excitation, that is $\Gamma x_1(t) \mathrm{d}W(t)$:
        \begin{equation}
            \left\{ 
            \begin{aligned}
                \mathrm{d}x_1(t) &= x_2(t) \mathrm{d} t\\
                \mathrm{d}x_2(t) &= -(2 \zeta x_2(t) + c_1 x_1(t)) \mathrm{d} t + \Gamma \mathrm{d}W(t)\\
            \end{aligned}
            \right.
            \label{eq:system_white}
        \end{equation}
        For the above system, the infinitesimal operator for a scalar function $F(x)$ becomes:
        \begin{equation}
            \begin{aligned}
                \mathcal{L} F(x(t)) = & x_2(t) \frac{\partial F(x(t))}{\partial x_1} \\
                & - (2\zeta x_2(t) + c_1 x_1(t)) \frac{\partial F(x(t))}{\partial x_2} 
            \end{aligned}
        \end{equation}
        Therefore, the additive Wiener process does not stabilize or destabilize the present system without parametric excitation. This is an important point to note.

        Concerning the system in Eq.~\eqref{eq:system_white}, several researches have been conducted in the 1960s-1970s~\cite{bogdanoff1962,kozin1963,caughey1965,Infante1968,Kozin1971}. Here, the authors briefly introduce these theories and numerical results.

        \subsection{Stability of second moment\cite{bogdanoff1962,kozin1986_2}}\label{subsec:stability_white_moment}
            In this subsection, the authors show the results based on the moment method~\cite{bogdanoff1962,kozin1986_2}. They apply the infinitesimal operator from 
            Eq.~\eqref{eq:infinitesimal_operator_use} to $x_1^2(t)$, $x_1(t)x_2(t)$, and $x_2^2(t)$, yielding:
            \begin{equation}
                \left\{
                \begin{aligned}
                    \mathcal{L}x_1^2(t)&=2 x_1(t) x_2(t)\\
                    \mathcal{L}x_1(t)x_2(t)&=x_2^2(t) - 2\zeta x_1(t)x_2(t) -c_1 x_1^2(t)\\
                    \mathcal{L}x_2^2(t)&=-4 \zeta x_2^2(t) - 2 c_1 x_1(t) x_2(t) + \Gamma^2 x_1(t)
                \end{aligned}
                \right.
            \end{equation}
            Here, the authors apply the following relation between the expectation and differentiation operators: 
            \begin{equation}
                \frac{\mathrm{d}}{\mathrm{d}t}\mathbb{E}[g(x(t))] = \mathbb{E}[\mathcal{L}g(x(t))].
            \end{equation}
            Then, the following moment equation can be derived:
            \begin{equation}
                \begin{aligned}
                    &\frac{\mathrm{d}}{\mathrm{d}t}
                    \begin{pmatrix}
                        \mathbb{E}[x_1^2(t)] \\
                        \mathbb{E}[x_1(t)x_2(t)] \\
                        \mathbb{E}[x_2^2(t)]
                    \end{pmatrix}
                    =\\
                    &\begin{pmatrix}
                        0 & 2 & 0 \\
                        -c_1 & -2 \zeta & 1 \\
                        \Gamma^2 & -2 c_1 & - 4 \zeta
                    \end{pmatrix}
                    \begin{pmatrix}
                        \mathbb{E}[x_1^2(t)] \\
                        \mathbb{E}[x_1(t)x_2(t)] \\
                        \mathbb{E}[x_2^2(t)]
                    \end{pmatrix}
                \end{aligned}
            \end{equation}
            This equation contains no information beyond the second-order moments and has a closed structure; however, as Bogdanoff and Kozin~\cite{bogdanoff1962} state, if a parametric excitation term of the colored noise is added, then the closed structure cannot be established anymore.

            For the above system, the characteristic equation becomes:
            \begin{equation}
                s^3 + 6 \zeta s^2 + (8 \zeta^2 + 4 c_1)s  + (8 c_1 \zeta - 2\Gamma^2) = 0
            \end{equation}
            Here, by applying the Routh--Hurwitz stability criteria, we can obtain an inequality that represents the stability boundary of the 2nd moment:
            \begin{equation}
                \Gamma^2 = 4 c_1 \zeta
            \end{equation}

        \subsection{Infante's approach\cite{Infante1968}}\label{subsec:stability_white_Infante}
        As the authors explained in section \ref{subsec:stability_color_Infante}, Infante’s result applies to the white noise case. Therefore, for our system, that is Eq.~\eqref{eq:system_white}, since $\mathbb[f(t)^2]=\sigma^2$, the boundary of stability is:
        \begin{equation}
            \Gamma^2 = 4 c_1 \zeta^2
        \end{equation}
        Kozin’s~\cite {kozin1973} results apply to Infante’s approach since $f(t)$ is not limited to white noise, allowing us to extend the results to the colored noise problem.
        
        \subsection{Kozin's approach\cite{kozin1973}}\label{subsec:stability_white_Kozin}
            Kozin~\cite{kozin1973} proposed a methodology to estimate a system's stability with multiplicative noise based on Khasminskii’s work~\cite{Khasminskii1967}. This methodology allows us to calculate $J_1$ to study a system's stability with colored noise, not just with white noise, as in Infante’s approach. 
            \begin{equation}
                \begin{aligned}
                    \mathcal{J}_1 = C \int_{-\pi/2}^{\pi/2} \biggl(&\cos 2 \varphi - \frac{4 \zeta}{\Gamma^2} \tan^2 \varphi\\
                    &+(1-c_1)\frac{2}{\Gamma^2}\tan\varphi\biggr) \eta_{\zeta}(\varphi) \mathrm{d} \varphi
                \end{aligned}
                \label{eq:def:Kozin_J1}
            \end{equation}
            Here, $C$ is a positive constant, and $\eta_{\zeta}$ is defined as follows:
            \begin{equation}
                \begin{aligned}
                    \eta_{\zeta}(\varphi) = & \exp \left[ -\frac{2}{3 \Gamma^2} \tan \varphi \left(3 c_1 + 3 \zeta \tan \varphi + \tan^2 \varphi \right)\right]\\
                    & \cdot \int_{-\pi/2}^{\varphi} \exp \biggl[ \frac{2}{3 \Gamma^2} \tan \theta \bigl(3 c_1 + 3 \zeta \tan \theta \\
                    & + \tan^2 \theta \bigr) \biggr] \sec^2 \theta \mathrm{d}\theta
                \end{aligned} 
            \end{equation}
            The integrand has singularities at $\pm \pi/2$, making its numerical computation challenging. Detailed descriptions can be found in~\cite{Kozin1971} and our previous literature~\cite{Maki2022stabilization}. Once $\mathcal{J}_1$ has been successfully calculated, the system's origin is considered stable if the following equation is satisfied.
            \begin{equation}
                \mathcal{J}_1 = 0
            \end{equation}

        \subsection{Arnold's approach for white-noise case\cite{Arnold1986Asymptotic}}\label{subsec:stability_white_Arnold}
            As shown in sec.~\ref{subsec:Arnold_Dostal}, Arnold obtained the stability boundary for the colored noise case, which the authors apply to the present white-noise problem. Now, they calculate eq.~\eqref{eq:def_spect_f}. In the present case, $C_f$ defined in eq.~\eqref{eq:def_Cf} becomes:
            \begin{equation}
                \begin{aligned}
                    C_f(t) &= \mathbb{E}[f(t)f(0)] = \delta(t),
                \end{aligned}
                \label{eq:def_Cf}
            \end{equation}
            where $\delta(t)$ denotes the Dirac’s delta function. Here, we define the spectral density of $C_f(t)$ as $S_{f_t f_t}$. Then, Eq.~\eqref{eq:Arnold_final_result} can be calculated.
            \begin{equation}
                \Gamma^2 = 8(1-\zeta^2)\zeta
            \end{equation}
            
        \subsection{Numerical results for white-noise system}
\begin{figure*}[tb]
                \centering 
                \includegraphics[width=1.0\hsize]{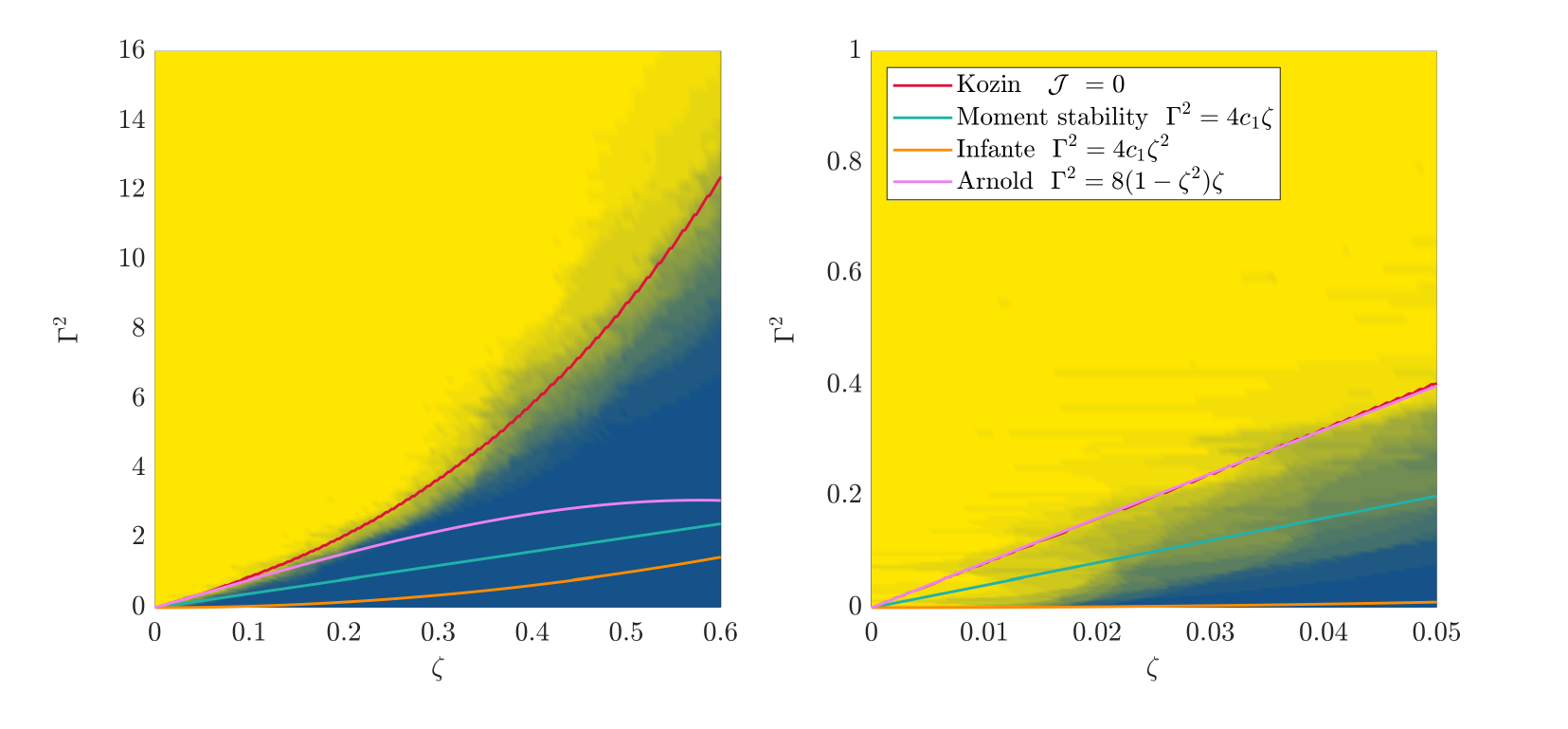}
                \caption{Comparison of the stability diagram between numerical and analytical results, $c_{1}=1.0$.}
                \label{fig:unstable_stabilized_c1_1}
            \end{figure*}

\begin{figure*}[tb]
                \centering 
                \includegraphics[width=1.0\hsize]{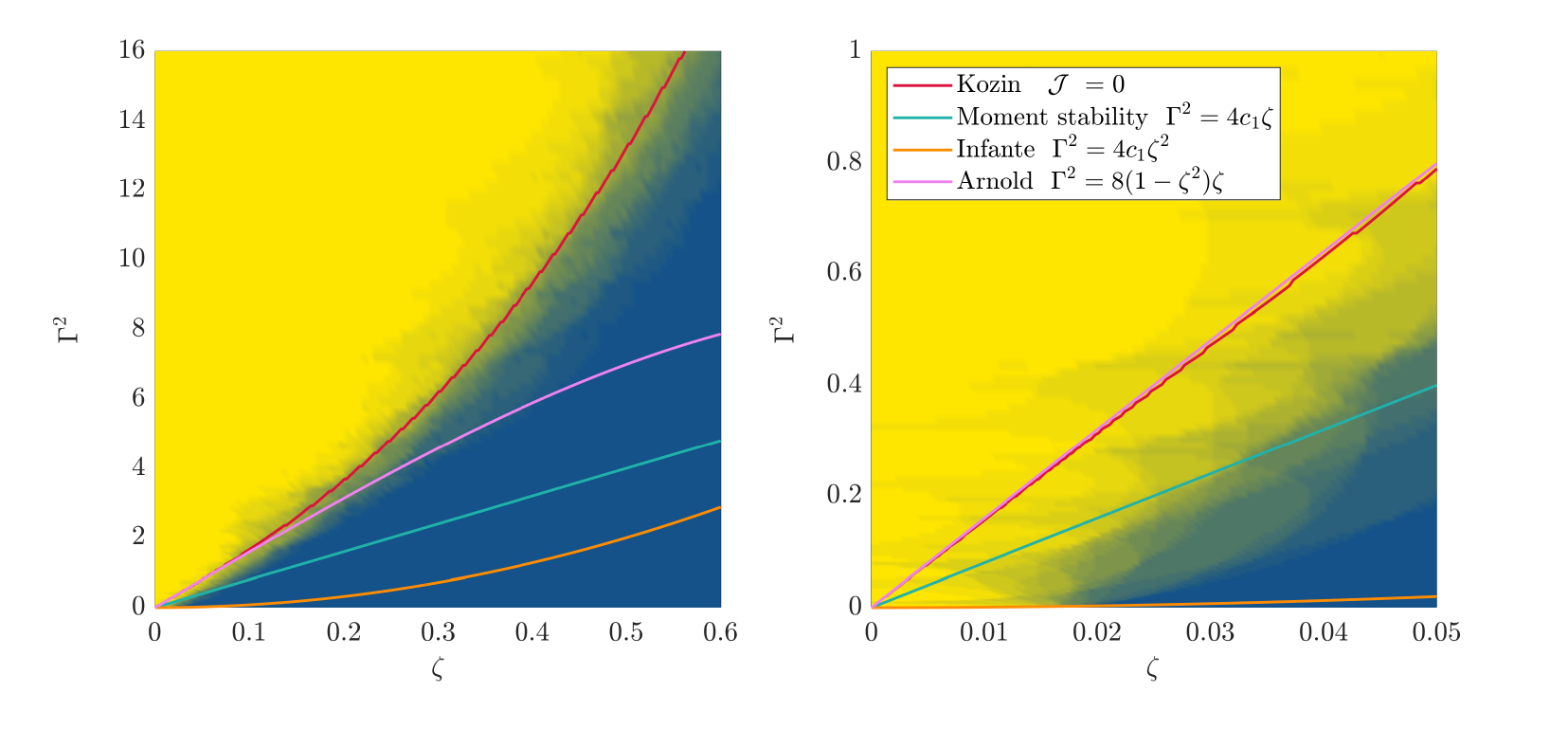}
                \caption{Comparison of the stability diagram between numerical and analytical results, $c_{1}=2.0$.}
                \label{fig:unstable_stabilized_c1_2}
            \end{figure*}

            Next, the authors show comparisons of the theoretical formulae with numerical results. A Monte Carlo simulation is employed using Euler Maruyama’s scheme~\cite{Klodenbook1992}. For each coefficient combination of $\zeta$ and $\Gamma$, 20 sample paths are generated, where the initial condition is $x_1(0) = 0.1$ and $x_2(0) = 0.1$. After a time of 160 sec, it is judged whether the path has converged to the system origin or not. 
            %
            Fig.~\ref{fig:unstable_stabilized_c1_1} and Fig.~\ref{fig:unstable_stabilized_c1_2} show the comparative results for $c_1 = 1$ and $c_1 = 2$, respectively. It can be seen that Kozin’s method accurately predicts the boundary of the stability. Additionally, Arnold’s method also predicts the boundary well in the vicinity of $\Gamma^2=0$, since it is based on the assumption of $\Gamma^2 \rightarrow 0$. However, the methods based on the 2nd moment stability and Infante’s method do not quantitatively correlate with MCS results, although the boundary trend can be qualitatively represented.

\section{Parametric oscillation for colored noise}\label{sec:colored_noise}
    %
    The stability of the following type of differential equation is analyzed in this section.
    \begin{equation}\label{eq:coloredparametric}
        \left\{\begin{aligned}
            \dot{x}_1(t) &= x_2(t)\\
            \dot{x}_2(t) &= - 2 \zeta x_2(t) - (c_1 +f(t) ) x_1(t) + h(t), 
        \end{aligned}\right.
    \end{equation}
    which includes the colored noise processes $f(t)$ and $h(t)$. 
    \subsection{Modeling of parametric excitation term}
         The following expressions of the parametric excitation terms are adopted in section \ref{sec:colored_noise}. 
        Let $f(t)$ be denoted as $P(t)$. The parametric excitation terms adopted in this section are calculated by first obtaining the $\mathrm{GM}$ variation, which is calculated from the computed restoring arm $\mathrm{GZ}$ for each regular wave height with a ratio of the ship length and wavelength of $1$ ($\lambda/L=1$) based on the Froude-Krylov assumption~\cite{Hamamoto1991}. This GM variation is then combined with the time series data of Grim’s effective waves to obtain the parametric excitation term $P(t)$.
        \begin{equation}
        \label{non-memory transformation}
            \displaystyle
            f(t) = P(t) = c_1 P^{\prime}(t).
        \end{equation}
        where $\omega_0 = \sqrt{c_1}$ or $c_1 = \omega_0^2$ denotes the natural roll frequency and $A_{\text{w}}$ denotes the effective wave amplitude. The parametric excitation terms in the section are expressed using the polynomial approximation of the relation between the change in the restoring force $\Delta \mathrm{GM}$ and wave amplitude amidships.

    Thereby, the processes $P(t)$ and $h(t)$ are stationary Gaussian processes with spectral density $S_{f_t f_t}$, and $S_{h_t h_t}$ as:
    
    \begin{equation}
        \left\{
        \begin{aligned}
            S_{f_t f_t} &= \frac{1}{2 \pi} \int_{-\infty}^{\infty} C_f(t) e^{-i \omega t} \mathrm{d}t\\
            S_{h_t h_t} &= \frac{1}{2 \pi} \int_{-\infty}^{\infty} C_h(t) e^{-i \omega t} \mathrm{d}t
        \end{aligned}
        \right.
        \label{eq:def_spect_f}
    \end{equation}
    where 
    \begin{equation}
        \left\{
        \begin{aligned}  
            C_f(t) &= \mathbb{E}[f(t)f(0)]\\
            C_h(t) &= \mathbb{E}[h(t)h(0)]
        \end{aligned}
        \right.
        \label{eq:def_Cf}
    \end{equation}
    It is worth noting that there exists a relation $G(\omega)=2S(\omega)$ between the two-sided spectrum $S(\omega)$ and the single-sided spectrum $G(\omega)$.
    \subsection{Infante's approach\cite{Infante1968}}\label{subsec:stability_color_Infante}
        In this subsection, the author slightly generalizes the method of Infante \cite{Infante1968} to the present problem. It is now assumed that the system is driven by physical noise (colored noise).
        \begin{equation}
            \ddot{x}_1(t) + 2 \zeta \dot{x}_1(t) + (c_1 + f(t))x_1(t)=0
            \label{eq:2nd_order_colored}
        \end{equation}
        Here, $f(t)$ is a zero-mean, stationary, ergodic physical noise. We rewrote the equation as follows:
        \begin{equation}
            \begin{gathered}
                \dot{x}(t) = [A + F(t)]x\\
                \left\{
                \begin{aligned}
                    A = \begin{bmatrix}
                    0 & 1\\
                    -c_1 & -2\zeta \\
                    \end{bmatrix},
                    F = \begin{bmatrix}
                    0 & 0\\
                    -f(t) & 0 \\
                    \end{bmatrix}
                \end{aligned}
                \right.
            \end{gathered}
        \end{equation}
        A positive-defined matrix $B$ is introduced as follows:
        \begin{equation}
            B = \begin{bmatrix}
            \alpha_1^2 + \alpha_2 & \alpha_1\\
            \alpha_1 & 1
            \end{bmatrix}
            \label{eq:def_Bmatrix}
        \end{equation}
        According to Infante’s paper~\cite{Infante1968}, if the following relation is satisfied for some positive $\epsilon$, then the main theory holds: let $\lambda_{\mathrm{max}}[X]$ be the maximum eigenvalue of matrix $X$.
        \begin{equation}
            \mathbb{E}[\lambda_{\mathrm{max}}[A^{\top}+F^{\top}(t)+B(A+F(t))B^{-1}]] < - \epsilon,
        \end{equation}
        then the system is almost asymptotically stable, yielding:
        \begin{equation}
            \begin{aligned}
                &\mathbb{E}[\lambda_{\mathrm{max}}[A^{\top}+F^{\top}(t)+B(A+F(t))B^{-1}]] \\
                &= -2 \zeta + \\
                &\sqrt{4(\zeta - \alpha_1)^2 + \frac{[\alpha_2 + \alpha_1^2 -c_1 - f(t) + 2 \alpha_1(\zeta-\alpha_1)]^2}{\alpha_2}}
            \end{aligned}
        \end{equation}
        Here, we define the elements of the $B$ matrix as follows: 
        \begin{equation}
            \left\{
            \begin{aligned}
                \alpha_1 &= \zeta\\
                \alpha_2 &= \zeta^2 + c_1
            \end{aligned}
            \right.,
        \end{equation}
        For the above, by using Schwarz’s inequality, the following result can be finally obtained:
        \begin{equation}
            \mathbb{E}[(f(t))^2] < 4 c_1 \zeta^2 
        \end{equation}
    \subsection{Arnold-Dostal's approach for colored-noise case\cite{Arnold1986Asymptotic,dostal2012non}}\label{subsec:Arnold_Dostal}
    Arnold et al.~\cite{Arnold1986Asymptotic} have obtained asymptotic results for the stability of the equation
    \begin{equation}\label{eq:linearrollequation3}
        \begin{aligned}
        \ddot{x}_1(t) + 2\zeta \dot{x}_1(t) + (c_1 + f(t))x_1(t) = 0,
        \end{aligned}
    \end{equation}
    where $f(t)$ denotes a stationary Gaussian process with spectral density $S_{f_t f_t}$.
    
    The top Lyapunov exponent was determined for the system of $y(t) = x_1(t) \exp(\zeta t)$:
    \begin{equation}
        \lambda_y=\lim_{t \rightarrow \infty} \log(|y(t)|^2+|\dot{y}(t)|^2)^{\frac{1}{2}},
    \end{equation}
    where $y(t)$ satisfies the following differential equation:
    \begin{equation}
        \begin{aligned}
            \ddot{y}(t) + [(c_1 - \zeta^2) - f(t)] y(t)  = 0,
        \end{aligned}
    \end{equation}

    The top Lyapunov exponent $\lambda_x$ was determined for the system of $x(t)$, which can be represented as:
    \begin{equation}\label{eq:lyapunovexpdamped}
        \begin{aligned}
            \lambda_x \approx -\zeta + \frac{\pi}{4(c_1-\zeta^2)}S_{f_t f_t}\left(2\sqrt{c_1 - \zeta^2}\right)\\
        \end{aligned}
    \end{equation}
    A negative Lyapunov exponent yields the stability of the corresponding system; thus, if $\lambda_x<0$ in Eq.~\eqref{eq:linearrollequation3}, the SDE in Eq.~\eqref{eq:linearrollequation3} is stable
    %
    The condition of $\lambda_x=0$ for negligibly small $P(t)$ becomes:
    \begin{equation}
        \begin{aligned}
            -\zeta + \frac{\pi}{4(c_1-\zeta^2)}S_{f_t f_t}\left(2\sqrt{c_1 - \zeta^2}\right) = 0.
        \end{aligned}
        \label{eq:Arnold_final_result}
    \end{equation}
    This is an implicit function for $\zeta$ or $c_1$ because $S_{f_{t}f_{t}}$  implicitly includes $H_{1/3}$. Therefore, iterative methods, such as Newton’s method, should be applied to obtain the stability boundary.

    \subsection{Ariaratnam and Tam's approach~\cite{Ariaratnam1979}}\label{subsec:results_of_ariaratonam}
    Ariaratnam and Tam~\cite{Ariaratnam1979} obtained the stability boundary with the use of the stochastic averaging method and utilized the outcome to Eq.~\eqref{eq:2nd_order_colored}.

    \color{black}
    Here, they analyze with the use of the stochastic averaging theorem proposed by Stratonovich\cite{Stratonovich1963} and Khasminskii\cite{Khasminskii1966}, the derived averaged equation for roll amplitude $A$ being as follows:
    \begin{equation}
        \begin{gathered}
            \mathrm{d}A = \left(- \alpha A + \frac{\beta}{2A}\right) \mathrm{d}t + \left(\gamma A^2 + \beta \right) \mathrm{d}W(t)\\
            \text{where}\,\,\,\left\{
            \begin{aligned}
                \alpha &= \zeta - \frac{3 \pi}{8 c_1}S_{f_t f_t}(\omega)\\
                \beta &= \frac{\pi}{c_1^2}S_{h_t h_t}(\omega)\\
                \gamma &= \frac{1}{4 c_1}S_{f_t f_t}(\omega)
            \end{aligned}
            \right.
        \end{gathered}
    \end{equation}
    Ariaratnam and Tam\cite{Ariaratnam1979} obtained the moment stability conditions for the equation they averaged.
    
    \subsubsection{First moment stability}
        \begin{equation}
            \zeta > \frac{3 \pi}{8 c_1} S_{f_t f_t}(2\omega_0)
        \end{equation}
    \subsubsection{Second moment stability}
        \begin{equation}
            \zeta > \frac{\pi}{2 c_1} S_{f_t f_t}(2\omega_0)
        \end{equation}
    \subsubsection{Condition on the PDF}
        By solving the Fokker-Planck equation, the stationary probability density function (PDF) was obtained as follows:
        \begin{equation}
            \begin{gathered}
                \mathcal{P}(A) = 2 \gamma \nu \beta^{\nu} \frac{A}{(\gamma A^2 + \beta)^{\nu + 1}}\\
                \text{where}\,\,\,\nu = \frac{1}{2} + \frac{\alpha}{\gamma}
            \end{gathered}
        \end{equation}
        For the above PDF, the condition of stability is considered to be $\nu > 0$, yielding:
        \begin{equation}
            \zeta > \frac{\pi}{4 c_1} S_{f_t f_t}(2\omega_0)
            \label{eq:Ariaratnam_Tam}
        \end{equation}
        The two results obtained here are identical to those obtained by Roberts \cite{Roberts1982parametric} in later years. These results also demonstrate that the external moment moment term, $h(t)$, does not affect the system's stability.

        Note that Eq.~\ref{eq:Ariaratnam_Tam} exactly matches the Arnold-Dostal result (Eq.~\ref{eq:Arnold_final_result}) if $\zeta$ is sufficiently small.
    

    %
    \subsection{Results of the energy-based averaging method \cite{dostal2012non,maruyama2022improved}}\label{subsec:results_of_energy_averaging}
\begin{figure}[tb]
        \centering 
        \includegraphics[width=0.8\hsize]{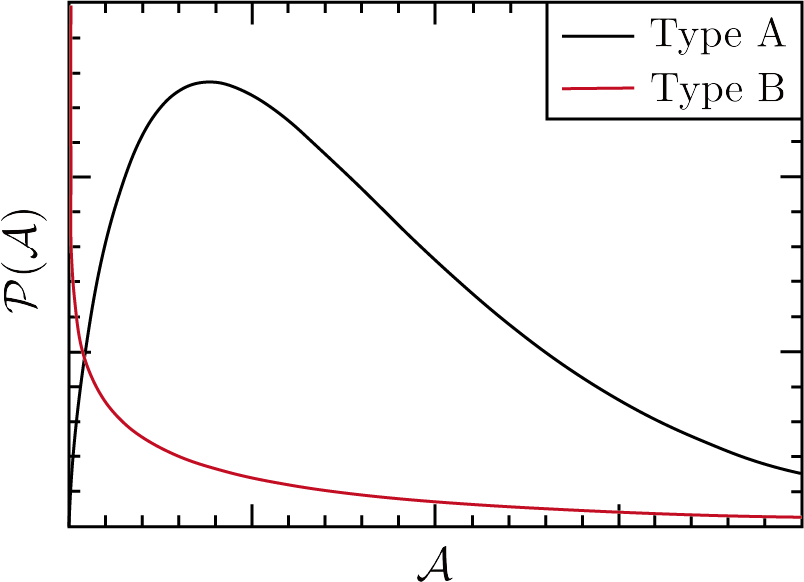}
        \caption{Schematic view of two PDFs}
        \label{fig:PDF_bifurcation}
    \end{figure}

    Here, we present an approach using an energy-based stochastic average method. The target system is as follows:
    \begin{equation}
        \ddot{x} + 2 \zeta\dot{x} + (c_1 + P(t)) x = 0
    \end{equation}
    Here, the Hamiltonian $\mathcal{H}$ of the system is:
    \begin{equation}
       \mathcal{H}(x, \dot{x}) = \dfrac{\dot{x}_{}^{2}}{2} + \dfrac{c_1}{2} x_{}^{2}.
    \end{equation}    
    Then, the following equation can be obtained:
    \begin{equation}
        \left\{\begin{array}{l}
        \dfrac{\mathrm{d}}{\mathrm{d}\,t} x=\dot{x} \\
        \\
        \dfrac{\mathrm{d}}{\mathrm{d}\,t} \mathcal{H}=- 2 \varepsilon \zeta \dot{x}^2-\sqrt{\varepsilon} P(t) x \dot{x}
        \end{array}\right.
    \end{equation}   
    The 1D SDE with respect to $\mathcal{H}$ can be obtained as:
    \begin{equation}
        \label{eq:SDE_hamiltonian}
       \mathrm{d}\mathcal{H} = ( m_{1}^{}(\mathcal{H}) + m_{2}^{}(\mathcal{H}) ) \mathrm{d}t + \sigma(\mathcal{H}) \mathrm{d}W
    \end{equation} 
    In the above equation, drift terms, i.e. $m_1$ and $m_2$, and diffusion term $\sigma^2$ can be represented by:
    \begin{equation}
        \left\{
        \begin{aligned}
            m_1^{}(\mathcal{H})&=-2 \zeta \mathcal{H}\\
            m_2(\mathcal{H})&=k \mathcal{H}\\
            \sigma^2(\mathcal{H})&=k \mathcal{H}^2
        \end{aligned}
        \right.
    \end{equation} 
    where
    \begin{equation}
       k=\dfrac{1}{c_1} \int_0^{\infty} R\left(\tau\right) \cos 2 \sqrt{c_1} \tau \mathrm{d} \tau
    \end{equation} 
    FPK equation can be obtained from Eq.(\ref{eq:SDE_hamiltonian}, and then the PDF for $\mathcal{H}$ is obtained as:
    \begin{equation}
       \begin{aligned}
           \mathcal{P}(\mathcal{H})&=\frac{C}{k \mathcal{H}^2} \exp \left(2 \int_{\mathcal{H}_{\text {imt }}}^\mathcal{H} \dfrac{-2 \zeta \theta+k \theta}{k \theta^2} \mathrm{d} \theta\right)\\
            &=C^{\prime} \mathcal{H}^{-\frac{4 \zeta}{k}}
       \end{aligned}
    \end{equation} 
    Here, $C$ and $C^{\prime}$ denote the normalization constant of the PDF.
    Using the transformation formula for the probability density, the probability density function for roll amplitude is:
    \begin{equation}
       \mathcal{P}(A)=\mathcal{P}(\mathcal{H}) \left|\dfrac{\mathrm{d}\mathcal{H}}{\mathrm{d}A}\right| =C^{\prime \prime} A^{\left(1-\frac{8 \zeta}{k}\right)} 
    \end{equation} 
    Here, $C^{\prime \prime}$ also denotes the normalization constant of the PDF.
    From this, the asymptotic behavior of the probability density function can be categorized into three cases for the relation between $k$ and the linear roll damping coefficient $2\zeta$.
    \begin{equation}
        \label{eq:parastab_hami}
        \begin{array}{l}
            \displaystyle
            (i)\quad\,\, 8 \zeta>k>0 \quad\, \lim_{A \rightarrow +0} \mathcal{P}(A) \rightarrow + \infty \\
            \\
            \displaystyle
            (ii)\quad\, 8 \zeta=k \quad \quad \,\,\,\,\,  \lim_{A \rightarrow +0} \mathcal{P}(A) = \mathrm{Const.} \\
            \\
            \displaystyle
            (iii)\quad k>8 \zeta>0 \quad \lim_{A \rightarrow +0} \mathcal{P}(A) \rightarrow 0
        \end{array}
    \end{equation} 
    Now calculate the equation of the boundary from $(ii)$ in Eq.(\ref{eq:parastab_hami}). Let $k$ denote the spectrum $S_{P}^{}(\omega)$ of $P(t)$. Therefore, we have:
    \begin{equation}
        k=\dfrac{\pi}{2 c_1}S_{P}^{}(2\sqrt{c_1})
    \end{equation} 
    By utilizing the above relation, $8 \zeta=k$ becomes as follows.
    \begin{equation}
        S_{P}^{}(2\sqrt{c_1})=\dfrac{16}{\pi}c_1\zeta .
    \end{equation} 
    However, even if $\mathcal{P}(A)\rightarrow +\infty$, it cannot be said that $\mathcal{P}(A)\neq 0$ when $A>0$, so it may be a non-conservative side estimation as shown in Fig.\ref{fig:parastab_C11}. This is because the condition in question represents the boundary between the behavior of two PDFs, Type A and Type B, as shown in Fig.\ref{fig:PDF_bifurcation}. This is the boundary of the bifurcation phenomenon of the probability density function; see, for example, page 506 of Arnold\cite{Arnold1995random}. Thus, this condition does not directly indicate the stability of the system's origin.

    Roberts \cite{Roberts1982parametric} presented a conditional expression for the probability density function from the FPK equation with $\lambda > 0$, where $\lambda$ is expressed as:
    \begin{equation}
        \lambda=\dfrac{\zeta-M}{M}
    \end{equation} 
    Here, $M$ means:
    \begin{equation}
        M=\dfrac{\pi}{8 c_1}S_{P}^{}(2\sqrt{c_1})
    \end{equation}
    Therefore, the boundary proposed by Roberts is:
    \begin{equation}
        S_{P}^{}(2\sqrt{c_1})=\dfrac{8}{\pi}c_1\zeta .
    \end{equation} 
     The above expression is obtained for the system with an external force. Furthermore, consider the following limit:
    \begin{equation}
        \lim_{A \rightarrow +0} \mathcal{P}(A) \rightarrow +\infty
    \end{equation} 
    By considering $1+2\lambda<0$, the following condition can be obtained: 
    \begin{equation}
        2\zeta<M
    \end{equation} 
    The boundary equation obtained from this method agrees with the boundary equation obtained from the energy-based stochastic averaging method.
    
    %
    \subsection{Results and discussions}\label{subsec:results_conclusion}
\begin{figure*}[tb]
        \centering 
        \includegraphics[scale=0.85]{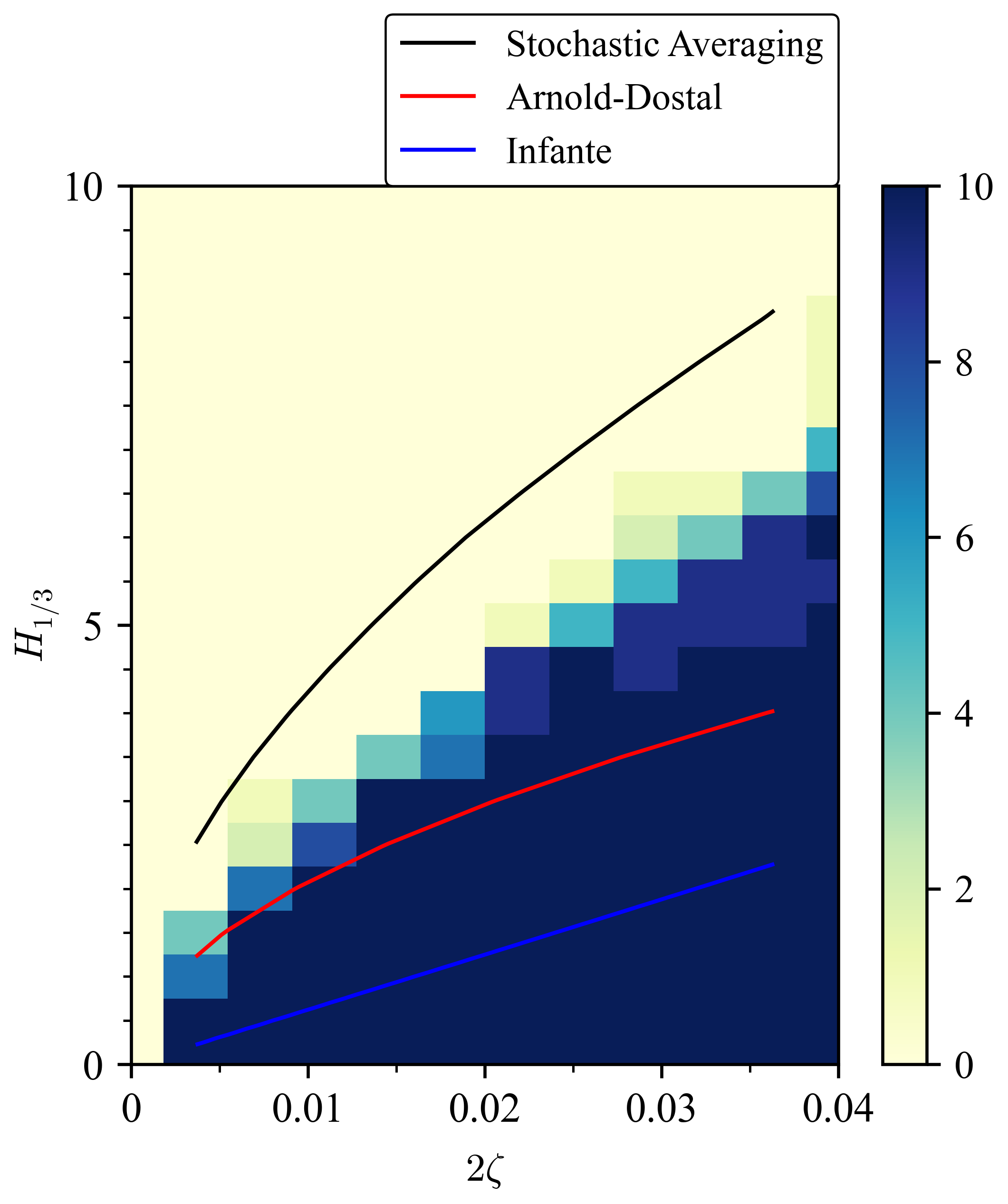}
        \includegraphics[scale=0.85]{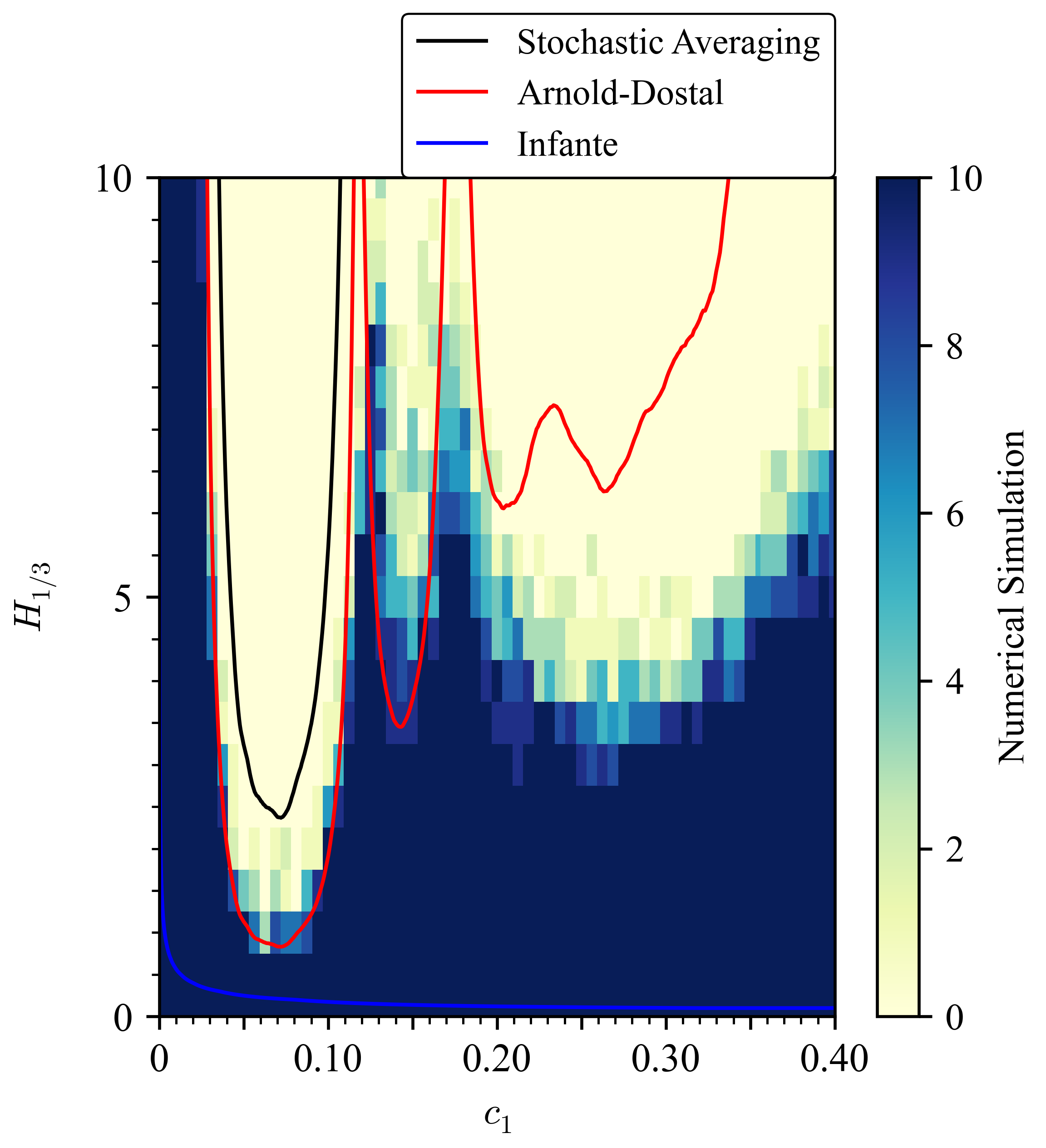}
        \caption{Comparison of the stability diagram between 'Arnold-Dostal's approach', 'Infante's approach', 'stochastic averaging method', and numerical simulation for C11 container ship parameters and forcing due to a ITTC spectrum with significant wave height $H_{1/3}$ and mean period $T_1=10$s.}
        \label{fig:parastab_C11}
    \end{figure*}

    This subsection compares the analytical conditions presented so far with numerical calculations.
    Fig.~\ref{fig:parastab_C11} shows the final results obtained for the ITTC spectrum. The subject ship is C11, and the graph shows a comparison of the results of theoretical calculations and numerical simulations of the stochastic averaging method, Infante's approach, and Arnold's approach. 



    \subsubsection{Infante's method}
    The results based on Infante's method show that the estimation is overly safe. This can be attributed to two factors. Firstly, Infante's method uses Eq.~\ref{eq:def_Bmatrix} to set the matrix $B$ related to the Lyapunov function. Even though the Lyapunov function is used to obtain the stability condition, it is a sufficient condition and may not be the optimal choice. This likely leads to an overly safe estimation. Secondly, Infante transforms the equation using Schwartz's inequality, another factor contributing to the overly safe estimation. 


    \subsubsection{Arnold-Dostal's method}
    From the figure, the results based on Arnold-Dostal's method explain the numerical results relatively well. However, since this theory is based on the assumption that $\sigma \rightarrow 0$, the estimation accuracy is expected to deteriorate as $\sigma$ increases, as seen in the results for white noise. Nonetheless, for the faced problem of vessel motion, the assumptions of Arnold-Dostal's method are considered to be valid to a great extent.
    
    \subsubsection{Averaging method}
    The analytical condition based on the averaging method appears to be essentially an estimation on the non-conservative side. This is because, as mentioned above, even if $\mathcal{P}(A) \rightarrow +0$ and $\mathcal{P}(A) \rightarrow +\infty$, $\mathcal{P}(A)=0$ is not necessarily guaranteed with $A>0$. Therefore, it is important to note that this is a condition that indicates the behavior of the probability density function near the origin, not directly indicating the origin's stability. Consequently, this method has the potential to be an estimation on the non-conservative side.


\section{Mitigation of parametric rolling due to rudder control}\label{sec:control}
    %
    
    As S{\"o}der et al.~\cite{soder2013parametric} have conducted, there is the potential to reduce the risk of parametric rolling using rudder control. The rudder is an inherent piece of equipment for the ship, making it an attractive option as it does not require any new active actuators or additional anti-rolling tanks. Furthermore, the control aims to restore the asymptotic stability of the system's origin, which is the major difference from regular roll mitigation controllers by rudder or fin-stabilizers. Thus, once the stability of the system's origin is restored, it is expected that roll motion will be completely eliminated in random head seas.
    \begin{equation}
        \ddot{x}_1(t) + 2 \zeta \dot{x}_1(t) + (c_1 + f(t))x_1(t) = f_{\mathrm{R}} \delta_{\mathrm{R}}
        \label{eq:system_white_rudder_original}
    \end{equation}
    Here, $f_R$ denotes the hydrodynamic derivative of the roll moment with respect to the rudder angle $\delta_{\mathrm{R}}$. Suppose that $\delta_{\mathrm{R}}$ has a feedback form as follows:

    \begin{equation}
        \delta_{\mathrm{R}}(t) = k_1 x_1(t) + k_2 x_2(t)
    \end{equation}
    In this research, the delay of rudder action is ignored for the sake of brevity; thereby, the system becomes:
    \begin{equation}
        \begin{gathered}
            \ddot{x}_1(t) + 2 \zeta^{\prime} \dot{x}_1(t) + (c_1^{\prime} + f(t))x_1(t) = 0\\
            \text{where}\,\left\{
                \begin{aligned}
                    2 \zeta^{\prime} &\equiv 2 \zeta - k_2\\
                    c_1^{\prime} &\equiv c_1 - k_1
                \end{aligned}
            \right.
        \end{gathered}
        \label{eq:system_white_rudder}
    \end{equation}
    If $2 \zeta^{\prime}$ and $c_1^{\prime}$, i.e., $k_1$ and $k_2$, are selected below the thresholds for stability, then parametric rolling can be completely prevented. The authors have demonstrated the control policy in the previous sections, and it can be said that control to increase the ``apparent damping force'' of $2 \zeta^{\prime} \equiv 2 \zeta - k_2$ will lead to complete prevention of parametric rolling.
    

\section{Conclusion}\label{sec:conclusion}
    In this study, the stability of the system with stochastically varying parametric excitation terms is discussed. Estimation formulae to predict the occurrence of instability because of the inclusion of multiplicative noise are introduced. The equations presented here are primarily based on those proposed by previous researchers, with only minor modifications to make them applicable to the parametric rolling of ships. The results of Arnold, for example, show that the stability boundaries can be captured with a relatively high accuracy, which is promising for practical use in the near future.

\begin{acknowledgements}
This study was supported by a Grant-in-Aid for Scientific Research from the Japan Society for the Promotion of Science (JSPS KAKENHI Grant \#22H01701). Further, this work was partly supported by the JASNAOE collaborative research program / financial support. The authors are also thankful to Enago (www.enago.jp) for reviewing the English language.
\end{acknowledgements}

%
\section*{Conflict of interest}

    The authors declare that they have no conflict of interest.

\bibliographystyle{spphys}       
\bibliography{main.bib}   

\end{document}